\documentstyle[11pt,newpasp,twoside,epsf]{article}
\def\edcomment#1{\iffalse\marginpar{\raggedright\sl#1\/}\else\relax\fi}
\reversemarginpar

\begin{document}
\title{Galactic nuclear activity induced by globular cluster merging}
 \author{R. Capuzzo--Dolcetta}
\affil{Dep. of Physics, Univ. "La Sapienza", P.le A.Moro 5, I-00185, Roma, Italy}

\begin{abstract}
The interpretation of the difference observed between the radial distribution of globular clusters and that of halo--bulge stars in elliptical galaxies is discussed in terms of evolution of their globular cluster systems. I present a short summary of the evidence that dynamical evolution of globular cluster systems is  not only able to explain  the flattening of their distribution toward the galactic center, but also it may have played an important role on the primordial activity of the parent galaxy.
\end{abstract}

\section{Introduction}
It is well known that the haloes and bulges of galaxies constitute \lq collisionless\rq~ systems, for its 2--body relaxation times are mich longer than their age. Globular clusters (GC), too, are halo objects but they are characterized by a mass $10^5-10^6$ heavier than the stars of the background where they orbit: this implies they do {\it not} constitute a collisionless system. 

The  collective effect of the fluctuating gravitational field produced by light stars along the cluster orbital motion reflects in the well known {\it dynamical friction} braking, that has been shown by various authors to be particularly important in triaxial galaxies, where the lack of symmetries in the potential excludes the conservation of any component of the angular momentum. This frictional braking acts on a time scale inversely proportional to the cluster mass and in triaxial galaxies it can be an order of magnitude shorter than in symmetric galaxies.
Moreover, for sufficiently massive clusters, the orbital energy is lost in less than 1 Gyr. In addition to the dynamical friction collective effect, on globular clusters acts also a {\it local} process: the {\it tidal} interaction with the (compact) galactic nucleus. The 2 evolutionary processes compete, in some sense, because dynamical friction tends to concentrate the GCS distribution around the galactic center, while a massive galactic nucleus (a super massive black hole?) may destroy a globular cluster which approaches too much to it, and this corresponds to an erosion of the GC inner galactic population.
\par\noindent 
The balance of the 2 mechanisms may lead to the explanation of the difference observed between the GCS and the halo-bulge stellar component radial distributions in some galaxies: the globular cluster systems (GCSs) are sistematically less concentrated to the parent galaxy center than bulges.
Another important consequence of the evolution of the GCS in a galaxy is the quantity of mass that, in form of stars tidally stripped by the nucleus, can feed the massive central black hole, releasing gravitational energy that can be radiated in form of electromagnetic waves, sustaining a violent nuclear activity. 
In this brief report we will describe and discuss some relevant aspects of this fascinating scenario, including that concerned with the stellar population of the inner kiloparsec of a galaxy.
\section{The globular cluster distribution in a galaxy}
The comparative discussion of the radial distributions of the stellar component and of globular clusters in elliptical galaxies has started not much time ago, due to the difficulty of determining in a reliable way the globular cluster distribution of the central galactic regions, even in nearby galaxies.
Grillmair et al. (1986) were the first to show how the stellar bulge in M 49 is more concentrated to the center of the galaxy than the GCS one. Much other evidence  of such phenomenon was later given (see, for instance, Lauer \& Kormendy 1986; Harris 1986; Harris et al. 1991), and now we can say that {\it no} case has been found yet where the GCS distribution in a galaxy is more peaked to its center than the surrounding halo-bulge stellar population.
\par The observational evidence does not have a certain interpretation.

\noindent Some claim that the reason of the different profiles is due to a difference in the formation epoch of the 2 halo components: globular clusters formed earlier, when the protogalaxy was more expanded, than the collisionless stellar component. 

\noindent Some others, starting from the reasonable hypothesis that the various components of the halo population formed almost simultaneously, think that the present difference is due to an {\it evolutionary} process. No clear evolutive processes can affect the stellar component as a whole, while the globular cluster distribution in a galaxy can suffer for both the collective effect of dynamical friction induced by stars and for the tidal, quasi impulsive, interaction with a compact nucleus sited in the galactic center. Shortly, dynamical friction is more effective on heavier clusters that are, usually, more tightly bound and so resist better to tidal shocks. As a consequence, the radial distribution of globular clusters in a galaxy should be mass (and density) dependent, provided the time scales of the mentioned effects are short enough in comparison with the age of the system.\\
We do not go here into details but just mention the work by Pesce, Capuzzo-Dolcetta \& Vietri (1992) and Capuzzo-Dolcetta (1993) who quantitatively showed how relevant are the frictional and tidal effects on the evolution of a GCS, particularly in a triaxial galaxy. Probably, the most relevant outcome of these works is that the time scales of the 2 evolutionary processes can be (in dependence on mass and density of clusters, indeed) short enough to be surely relevant for the evolution of GCSs from their initial state.
\par A straightforward application of the results of Pesce et al. (1992) and Capuzzo-Dolcetta (1993) lead Capuzzo-Dolcetta \& Tesseri (1997) to a succesful comparison of the evolved theoretical GCS spatial distributions (at varying some relevant parameters of the model) with the general characteristics of the observed distributions like, for example, the observed larger core radii of  the GCS radial distributions with respect to the bulge one. So, the evolution of the GCS in an elliptical galaxy seems a confirmed hypothesis, and, other explaining the mentioned difference in radial profiles, it may have played an important role in the primeval lives of galaxies, supplying a significant amount of matter to the central galactic regions, often sites of violent activity (AGNs).

\subsection{The comparison between the GCS and stellar bulge radial distributions}
In addition to our galaxy, M 31 and M 87 (discussed in Capuzzo-Dolcetta \& Vignola 1997) the comparison between the stellar bulge and globular cluster radial distributions has been extended in details to a set of 14 elliptical galaxies observed with the WFPC of the HST by Forbes et al. (1996), Forbes et al. (1998a,b).
\begin{table}
\center
\caption{The presently observed number of  clusters  ($N$), its initial value ($N_i$), its fractional variation ($\Delta$), the mass lost in form of disappeared globulars ($M_l$). Data are from Capuzzo--Dolcetta \& Vignola (1997), Capuzzo--Dolcetta \& Tesseri (1999) and Capuzzo--Dolcetta \& Donnarumma (2000).}
\label{symbols}
\begin{tabular}{@{}lllll}
\\ Galaxy & $N$ & $N_i$ & $\Delta $ & $M_l$ ($M_{\odot}$) \\
\tableline
Milky way & 155 & 211 & 0.26 & $1.80 \times 10^7$\\
M 31 & 283 & 368 & 0.23 & $2.30 \times 10^7$\\
M 87 & 4456 & 8021 & 0.44 & $2.33 \times 10^9$\\
  NGC 1379 & 132 & 512 & 0.74 & $1.50 \times 10^8$\\
  NGC 1399 & 5168 & 9680 & 0.63 & $1.44 \times 10^8$\\
  NGC 1404 & 508 & 1061 & 0.53 & $1.75 \times 10^8$\\
  NGC 1427 & 248 & 487 & 0.49 &  8.86  $\times 10^7$\\
  NGC 1439 & 130 & 141: & 0.08: &  4.79 $\times 10^6:$\\
  NGC 1700 &  25 & 39: & 0.36: &   3.66  $\times 10^6:$\\
  NGC 4365 & 517 & 849 & 0.39 &  $7.48 \times 10^7$\\
  NGC 4494 & 200 & 297 & 0.33 &  2.98 $\times 10^7$\\
  NGC 4589 & 241 & 371 & 0.35 &   7.58  $\times 10^7$\\
  NGC 5322 & 175 & 266 & 0.34 &   6.51  $\times 10^7$\\
  NGC 5813 & 382 & 596 & 0.36 &  1.03 $\times 10^8$\\
  NGC 5982 & 135 & 260 & 0.48 &   8.86  $\times 10^7$\\
  NGC 7626 & 215 & 365 & 0.41 &   3.59 $\times 10^8$  \\
  IC 1459  & 271 & 516 & 0.47 &  1.57 $\times 10^8$\\
\tableline
\tableline
\end{tabular}
\end{table}
\noindent
The results, presented in Capuzzo-Dolcetta \& Tesseri (1999) and Capuzzo-Dolcetta \& Donnarumma (2000), show in all the cases a significantly flatter distribution of the GCS. \\
\noindent Assuming that initially the stellar bulge and the GCS had the same profile it is easy to evaluate the number of globular clusters \lq lost \rq~ in a galaxy as the difference of the surface integrals of the 2 profiles. A proper estimate of the mean value of the globular cluster initial mass brings to a value of the quantity of mass lost to the central galactic region. This was done in the mentioned papers;
as a summary of the results, in Table 1 we give the values relative to the set of 17 galaxies studied so far. It is really remarkable the high percentage, 41\%, of the initial GCS mass that, according to our estimates, is fallen, in the average, in the inner galactic zones. The quantity of mass lost to the central regions, averaged over the set of 17 galaxies, is $2.19\times 10^8$ M$_\odot$, that should have been relevant as a source of fuel for a possible central compact object.

\section{Nucleus feeding by centrally  merged globular clusters}
As shown in Capuzzo-Dolcetta (1993) the possible causes of the significant difference observed between the stellar bulge and the GCS distribution in a galaxy are dynamical friction and tidal disruption, this latter induced by a compact galactic nucleus, if present. In that paper a preliminary quantitative discussion of the modes of formation of a supercluster composed by orbitally decayed clusters merged at the galactic center, as well as of debris of tidally shattered ones has been discussed and demonstrated to occur in times acceptably short to give birth to a  supermassive object which could release stars to the central black hole in a way to make it a powerful engine of electro--magnetic energy extracted by the gravitational field. A subsequent, still in preparation, paper by Capuzzo-Dolcetta (2001) presents a detailed quantitative model of the evolution of a GCS in a triaxial galaxy leading to a detailed estimate of the time growth of the central compact seed as well as of the luminosity evolution of the resulting AGN. Of course, the model depends on some parameters like, mainly, the initial velocity dispersion of globular clusters, their IMF, their orbital distribution, etc.; these parameters can just partially be fixed a priori. By the way, it is relevant noting that a limitation of the possible values of these parameters is induced by the request that the models give, at the present time, acceptable numbers of observable clusters with the proper observed radial distribution and velocity dispersions, in order to be fully  self-consistent.\\
We do not go here into details of the models, but rather want just to give a qualitative insight to the phenomenology, starting from first principles, that is useful to convince that the overall scenario is realistic at all.\\
Suppose that around a central galactic black hole (b.h.) of mass $m_{bh}$ there is an almost steady distribution of stars with typical mass and radius $m_*$ and $R_*$, and having space density $\rho_*$ and average velocity $<v_*>$, then the rate of mass accretion onto the b.h. is simply $\dot m = \rho_* \sigma <v_*>$.
In the mass accretion rate  $\sigma$ is the \lq capture \rq~ cross--section given by
$$
\sigma = \pi r_d^2\left({  1+{ {Gm_{bh}\over r_d} 
\over { {1\over 2} <v_*^2 >}} }\right)=
\pi r_d^2\left({ 1+2{r_{inf}\over r_d} }\right),
$$
where $r_{inf}$ is the \lq influence \rq~ radius of the b.h. and $r_d$ is the \lq destruction \rq~ radius (below this distance from the b.h. the star is tidally destroyed or swallowed) given by
$$r_d = max(r_t,r_S),$$
where  $r_t$ is the  \lq tidal \rq~ radius and $r_S$ the Schwarzschild's radius of the b.h., respectively. Because the ratio
$$
{r_S\over r_t}={2^{2\over 3} G \over c^2} {m_*^{1\over 3} \over R_*}
m_{bh}^{2\over 3} 
$$
is $\geq 1$ just for massive b.h. ($m_{bh} \geq m_{bh,crit} \simeq 1.62 \times 10^8$ M$_\odot$ assuming solar values for the cluster star mass and radius) it is clear that the process of b.h. feeding is by mean of  tidally disrupted stars until the b.h. mass has grown enough to swallow the surrounding stars whole when entering the  event horizon through the Schwarzschild's radius. In the 2 cases we have (assuming $<v_*^2>=<v_*>^2$) 

$$
\dot m (M_\odot yr^{-1})= \left\{ \begin{tabular}{ll}
$8.2\times 10^{-16} m_{bh}^{4/3}m_*^{-1/3}\rho_* R_*<v_*>^{-1}$, & $r_d=r_t$,\\
 $2.8\times 10^{-21} m_{bh}^2\rho_* <v_*>^{-1}$ & $r_d=r_S$,
\end{tabular}\right.
$$

\noindent respectively ($m_*$, $m_{bh}$, $\rho_*$ and $R_*$ are in solar units and $<v_*>$ in km s$^{-1}$). Let us note that, of course, the efficiency of conversion into radiation is higher in the case when stars are tidally destroyed before falling onto the compact object than when swallowed whole.\\
On the basis of previous equations, it is possible to draw the plot given in Fig. 1a, where, in function of $m_{bh}$, the minimum values of $\rho_* <v_*>^{-1}$ required to yield an accretion rate of at least 1 M$_\odot$ yr$^{-1}$ are given (1 M$_\odot$ yr$^{-1}$ is the typical value of mass accretion required to sustain an AGN) yielding a separation into a \lq high\rq~ accretion and a \lq low\rq~ accretion density zone.

\begin{figure}
\plottwo{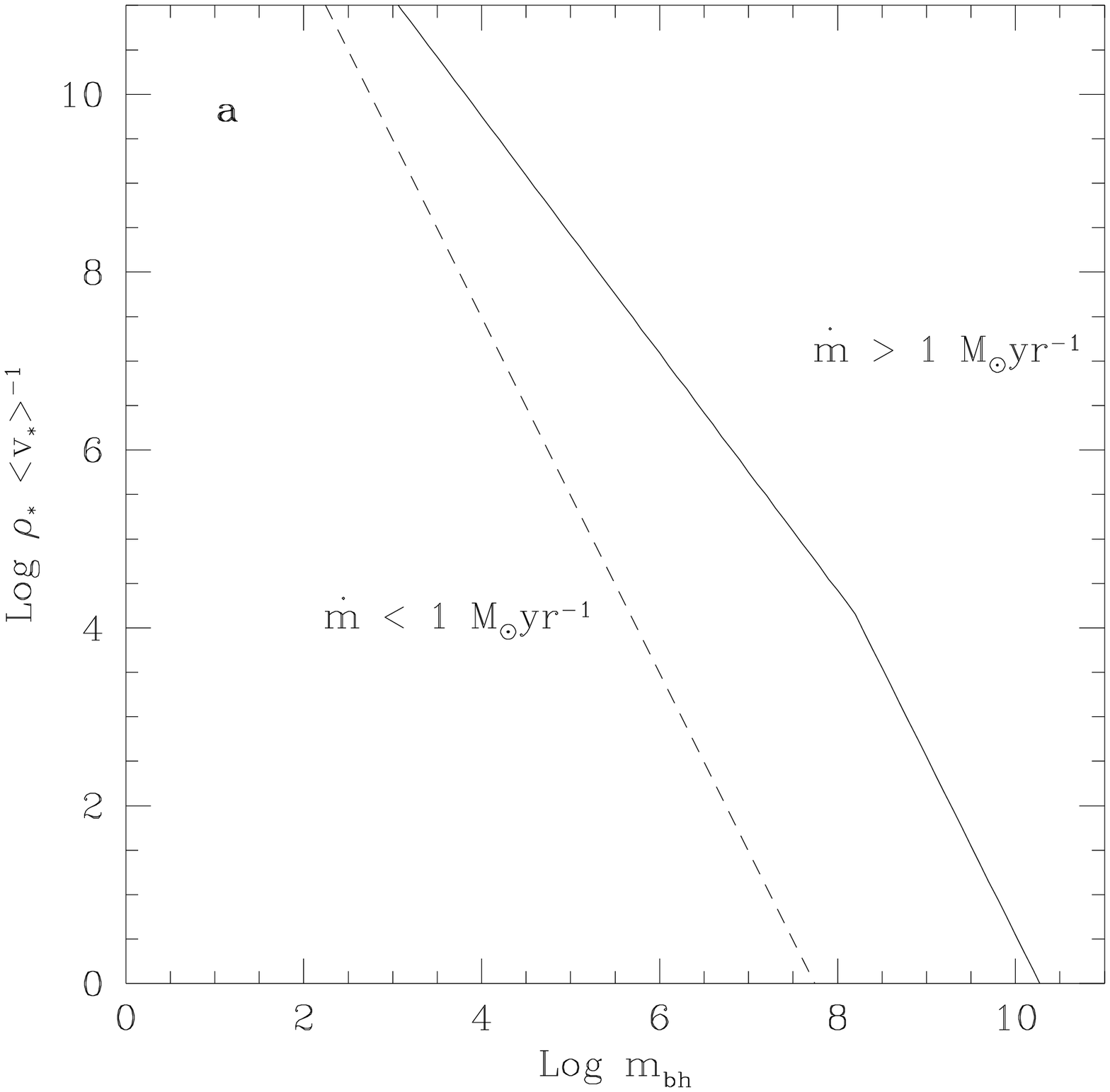}{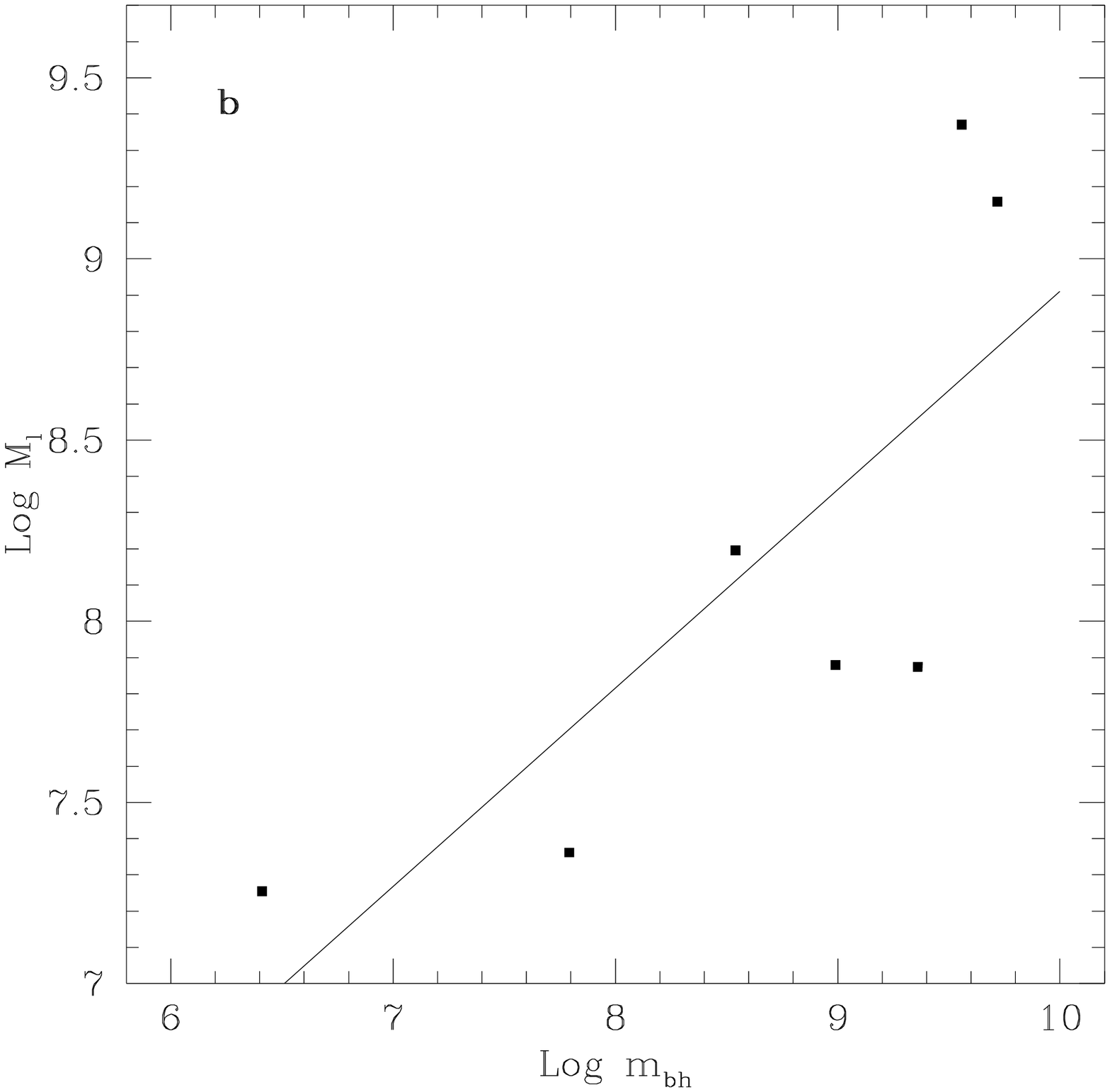}
\caption{panel {\bf a}: Bi-logarithmic plot of the curve (solid line)\\ $\rho_*<v_*>^{-1}$
(in units M$_\odot$ pc$^{-3}$/(kms$^{-1}$)) that corresponds to $\dot m = 1$ M$_\odot$ yr$^{-1}$ in function of $m_{bh}$.  The dashed line is the \lq collisional \rq~ curve (see text).\\
panel {\bf b}: The correlation between the (logarithmic) GCS mass lost and the central galactic black hole  mass for 7 galaxies for which these data are available. Masses are in solar masses.
The  straight solid line is the least square fit to the data.} 
\end{figure}
As an example, the known values of the central b.h. masses in our galaxy, M31 and M 87 ($2.6\times 10^6$, $6.2\times 10^7$, $3.6\times 10^9 M_\odot$, respectively) imply a stellar density  at least $4 \times 10^7$ M$_\odot$, $6.3 \times 10^5$ M$_\odot$, $400$ M$_\odot$, respectively, to accrete  $1$ M$_\odot$ yr$^{-1}$ in the hypothesis of $<v_*> = 10$ km s$^{-1}$ (which is, anyway, just a typical value of the mean stellar velocity in a globular cluster). It is remarkable that these density values are not too high to be reached by a stellar system composed by merged globulars, at least in its central region (many galactic globulars have central densities greater than $ 10^4$ M$_\odot$ pc$^{-3}$, and few of them greater than $ 10^5$ M$_\odot$ pc$^{-3}$).
It may be interesting noting that in particular situations of very high dense clusters surrounding the compact nucleus, the accretion can occur on the larger space scale given by the $collisional$ radius
$$
r_{coll}={Gm_{bh} \over v_{e*}^2}= {1\over 2} {m_{bh}\over m_*} R_*,
$$
($v_{e*}$ is the escape velocity from star surface)
which determines the radius of the sphere centred on the b.h.  where star-star physical collisions are disruptive  (see e.g. Binney and Tremaine 1998). Actually, 
it turns out that in normal conditions $r_{coll}$ is the largest of the 3 radii ($r_t$, $r_S$, $r_{coll}$), being 

$$
{r_S\over r_{coll}} = 2 {r_{S*}\over R_*} <<1,
$$

being $r_{S*}$ is the star Schwarzschild radius, and 

$$
{r_t\over r_{coll}} =2^{4/3} \left({ {m_*\over m_{bh}} }\right)^{2/3}\leq 1
$$
whenever $m_{bh}/m_*\geq 4$.\\
 A larger $r_d$ corresponds,  of course, to a larger $\sigma$ and so it $would$ allow a huge mass accretion onto the b.h., as given by
$$
\dot m_{coll}(M_\odot yr^{-1})=3.27\times 10^{-16} m_{bh}^2m_*^{-1}\rho_* R_*<v_*>^{-1}.
$$
For the sake of comparison,  the line corresponding to this 
$physically$ $collisional$ regime is also shown in Fig. 1a.
\par
To conclude, it is a relevant point in favour of the GCS evolutionary picture the existence of a positive correlation (as shown in Figure 1b) among the values of the (possible) mass lost in form of orbitally decayed clusters ($M_l$ in Table 1) and those of the central b.h. in the 7 galaxies for which these data are available (the least square straight line fitting the whole data set has a correlation factor $r^2=0.61$ and standard error $=0.56$). Actually, as remarked in Capuzzo--Dolcetta \& Tesseri (1999) a positive correlation is expected because larger black holes are grown during the evolution and they should induce a greater evolution of the GCS.

\section{Conclusions}
I have presented a summary of some results concerning the globular cluster distribution in galaxies as compared with that of the underlying stellar component.

 It is known that globular clusters have a less peaked distribution; if this centrally flatter behaviour is interpreted in evolutive terms, it may imply a number of \lq missing \rq~ clusters that can be estimated in the assumption of an initial cluster distribution equal to that of stars of the halo-bulge. The found number of globular clusters likely lost to the inner galactic region is  evaluated to be significant in the galaxies studied so far. This may have had important consequences on the initial life of the galaxy, for it corresponds to a large quantity of mass in form of stars that have been buzzing around the center of the galaxy feeding a compact nucleus therein and allowing it to emit energy at the levels of AGNs. Another consequence of the globular cluster system evolution in a galaxy is the contribution to the inner field, in a way that has not been yet deeply analysed.\\
Rather than going into details of the model describing the evolution of the coupled globular clusters-central black hole system in an elliptical galaxy (that can be found in other longer papers referred in this report), in this short paper I preferred to give simple evaluations of the densities required by the cluster around the galactic central black hole in order to be able to release it a sufficient quantity of mass to feed it as engine of an AGN.

\end{document}